\documentclass[aps,prd,nofootinbib,preprint,tightenlines,showpacs]{revtex4}
\newcommand{\be}{\begin{equation}}
\newcommand{\ee}{\end{equation}}
\newcommand{\bea}{\begin{eqnarray}}
\newcommand{\eea}{\end{eqnarray}}
\def\bml{\begin{subequations}}
\def\eml{\end{subequations}}
\def\blea{\bml\bea}
\def\elea{\eea\eml}
\def\bx{\mathbf{x}}
\def\Rmax{R_{\text{max}}}
\def\sgn{\mathop{\rm sgn}}
\def\Tsplit{T^{\text{split}}}
\def\Tren{T^{\text{ren}}}
\usepackage{amsmath}
\usepackage{amsfonts}
\usepackage{latexsym}

\begin{document}

\title{Quantum inequality in spacetimes with small curvature}

\author{Eleni-Alexandra Kontou}
\author{Ken D. Olum}
\affiliation{Institute of Cosmology, Department of Physics and Astronomy,\\ 
Tufts University, Medford, MA 02155, USA}

\begin{abstract}
Quantum inequalities bound the extent to which weighted time averages
of the renormalized energy density of a quantum field can be
negative. They have mostly been proved in flat spacetime, but we need
curved-spacetime inequalities to disprove the existence of exotic
phenomena, such as closed timelike curves. In this work we derive such
an inequality for a minimally-coupled scalar field on a geodesic in a
spacetime with small curvature, working to first order in the Ricci
tensor and its derivatives.  Since only the Ricci tensor enters, there
are no first-order corrections to the flat-space quantum inequalities
on paths which do not encounter any matter or energy.

\end{abstract}

\pacs{04.20.Gz 
      03.70.+k 
}

\maketitle

\section{Introduction}
\label{sec:intro}

In the context of General Relativity all kinds of exotic spacetimes
are allowed. With the appropriate stress-energy tensor $T_{\mu \nu}$,
following Einstein's equations, the spacetime can contain wormholes
and allow superluminal travel and the construction of ``time
machines''. However, in quantum field theory, there are restrictions
on $T_{\mu \nu}$.  Two examples of these are the energy conditions and
the quantum inequalities.  Pointwise energy conditions bound the
stress-energy tensor at each spacetime point, but they are easily
violated, since quantum field theory allows arbitrary negative
energies (e.g., in the Casimir effect).  On the other hand, averaged
energy conditions bound the stress-energy tensor integrated along a
complete geodesic and quantum inequalities bound a weighted time
average of the total energy. These have been proven to hold in a
variety of spacetimes.

Ford \cite{Ford:1978qya} introduced quantum inequalities to prevent
the violation of the second law of thermodynamics. After that, quantum
inequalities were derived for various spacetimes and fields. The
majority of these results are for free fields on flat spacetimes
without boundaries, while a few are for interacting fields in
spacetimes with less than four dimensions
\cite{math-ph/0412028,arXiv:1304.7682}. For spacetimes with boundaries
there are difference quantum inequalities, which bound the difference
of $T_{\mu \nu}$ between some state and a reference state.  But these
inequalities cannot be used to rule out exotic spacetimes arising from
vacuum energies.

Energy conditions have been used to address the possibility of exotic
spacetimes. Specifically, Ref.~\cite{Graham:2007va} showed that the
achronal averaged null energy condition (achronal ANEC) is sufficient
to rule out most known spacetimes with exotic curvature. In previous
work \cite{Kontou:2012ve}, we proved achronal ANEC for spacetimes with
a classical source.  However, to do that we assumed that with a
timescale small compared to any curvature radius the quantum
inequality for flat spacetime still holds with small corrections. Ford,
Pfenning and Roman \cite{Ford:1995wg, Pfenning:1997wh} also have
suggested that the flat-space quantum inequalities can be used in
spacetimes with small curvature. However none of these results have
been explicitly proven.

Fewster and Smith \cite{Fewster:2007rh} proved an absolute quantum
inequality (i.e., one without dependence on a reference state) that
applies to spacetimes with curvature. Their bound involves the Fourier
transform of differentiated terms of the Hadamard series up to fifth
order. In recent work \cite{Kontou:2014eka}, we used their result to
provide a bound for flat spacetimes with a background potential. In
the same paper we also showed that is sufficient to consider only
terms up to first order, which makes Fewster and Smith's result more
practical. Using this result we will now show, in accordance with our
past conjecture and previous work, that in spacetimes with small
curvature, the quantum inequality is the same as in flat space with
small corrections that depend on the curvature.

The present paper closely follows Ref.~\cite{Kontou:2014eka}.  We
begin by stating the general absolute quantum inequality of Fewster
and Smith \cite{Fewster:2007rh} in Sec.~\ref{sec:FSQEI}. The
inequality bounds the time averaged, renormalized energy density using
the Fourier transform of a point-split energy operator applied to
$\tilde{H}$, which is a combination of the Hadamard series and the
advanced-minus-retarded Green's function. In Sec~\ref{sec:tsplit} we
discuss and simplify this operator. In Sec.~\ref{sec:iE} we compute
the Green's function to first order for a spacetime with curvature,
and in Sec.~\ref{sec:H} we use that result to calculate
$\tilde{H}$. In section \ref{sec:tsplitH} we apply the point-split
energy operator and compute $\tilde{H}$. Finally we perform the
Fourier transform, and find the resulting quantum inequality in
Sec.~\ref{sec:QI}.  We conclude in Sec.~\ref{sec:conclusions}.

We use the sign convention $(-,-,-)$ in the classification of Misner,
Thorne and Wheeler \cite{MTW} . Indices $a,b,c, \dots$ denote all
spacetime coordinates while $i,j,k \dots$ only spatial coordinates.

\section{Absolute Quantum Energy Inequality}
\label{sec:FSQEI}

We consider a massless, minimally-coupled scalar field with the usual
classical stress-energy tensor,
\be
T_{ab}=\nabla_a \Phi \nabla_b \Phi-\frac{1}{2} g_{ab} g^{cd} \nabla_c
\Phi \nabla_d \Phi \,.
\ee
Let $\gamma$ be any timelike geodesic parametrized by proper time $t$,
and let $g(t)$ be any any smooth, positive, compactly-supported
sampling function.  In flat spacetime, Fewster and Eveson
\cite{Fewster:1998pu} showed that
\be\label{flatQI}
\int_{-\infty}^\infty dt \, T_{tt}(\gamma(t)) g(t)^2 \ge
-\frac{1}{16\pi^2}\int_{-\infty}^\infty dt \,  g''(t)^2 \,.
\ee
We will generalize Eq.~(\ref{flatQI}) to geodesics in curved
spacetime.

First we construct Fermi normal coordinates \cite{Manasse:1963zz} in
the usual way: We let the vector $e_0(t)$ be the unit tangent to the
geodesic $\gamma$, and construct a tetrad by choosing arbitrary
normalized vectors $e_i(0), i=1,2,3$, orthogonal to $e_0(0)$ and to
each other, and define $\{e_i(t)\}$ by parallel transport along
$\gamma$.  The point with coordinates $(x^0,x^1,x^2,x^3)$ is found by
traveling unit distance along the geodesic given by $x^i e_i(x^0)$
from the point $\gamma(x^0)$.

We work only in first order in the curvature and its derivatives, but
don't otherwise assume that it is small.  We assume that the
components of the Ricci tensor in any Fermi coordinate system, and
their derivatives, are bounded,
\bea \label{Rmax}
|R_{ab}| \leq \Rmax \qquad |R_{ab,cd}| \leq \Rmax'' \qquad |R_{ab,cde}| \leq \Rmax'''\,.
\eea
Eqs.~(\ref{Rmax}) are intended as universal bounds which hold without
regard to the specific choice of Fermi coordinate system above.  We
will not need a bound on the first derivative.  The reason that we
bound the Ricci tensor and not the Riemann tensor is that, as we will
prove, the additional terms of the quantum inequality do not depend on
any other components of the Riemann tensor. We will discuss this
result further in the conclusions.

Following Ref.\cite{Fewster:2007rh}, we define the renormalized energy
density
\be \label{Tren}
\langle \Tren_{tt} \rangle \equiv \lim_{x\to x'}  \Tsplit \left( \langle \phi(x)\phi(x') \rangle-H(x,x') \right)-Q+C_{tt}\,,
\ee
with quantities appearing in Eq.~(\ref{Tren}) defined as follows. $\Tsplit$ is  the point-split energy density operator,
\be \label{tsplit}
\Tsplit=\frac{1}{2}\sum_{a=0}^3 e^\alpha_a \nabla_\alpha \otimes
 e_a^{\beta'} \nabla_{\beta'}=\frac{1}{2}\sum_{a=0}^3 \partial_a
 \partial_{a'}\,.
\ee
where $\partial_a f$ or $f_{,a}$ denotes the gradient of a function
$f$ with respect to $x$ in the direction of $e_a(x)$, and
$\partial_{a'} f$ or $f_{,a'}$ the same with $x'$ in place of $x$.

We renormalize the energy density according to the procedure of Wald
\cite{Wald:qft}, by taking the difference between the two point
function, $\langle \phi(x)\phi(x') \rangle$, and the Hadamard series,
\be \label{hadamard}
H(x,x')=\frac{1}{4\pi^2} \left[ \frac{\Delta^{1/2}}{\sigma_+(x,x')}+\sum_{j=0}^{\infty}v_j(x,x') \sigma_+^j (x,x') \ln(\sigma_+(x,x'))+\sum_{j=0}^{\infty}w_j (x,x')\sigma^j (x,x') \right] \,,
\ee
where $\sigma$ is the squared invariant length of the geodesic between
$x$ and $x'$, negative for timelike distance. In flat space
\be
\sigma(x,x')=-\eta_{ab} (x-x')^a (x-x')^b \,.
\ee
By $F(\sigma_+)$, for some function $F$, we mean the distributional
limit
\be 
F(\sigma_+)=\lim_{\epsilon \to 0^+} F(\sigma_{\epsilon}) \,,
\ee
where
\be
\sigma_{\epsilon}(x,x')=\sigma(x,x')+2i \epsilon(t(x)-t(x'))+\epsilon^2 \,.
\ee
In some parts of the calculation it is possible to assume that both
points lie on the geodesic, so we define
\be
\tau=t-t'
\ee
and write
\be
F(\sigma_+)=F(\tau_-)=\lim_{\epsilon \to 0} F(\tau_\epsilon) \,,
\ee
where
\be
\tau_\epsilon=\tau-i\epsilon \,.
\ee
The function $\Delta$ is the van Vleck-Morette determinant bi-scalar, given by
\be \label{delta}
\Delta(x,x')=-\frac{\det (-\nabla_a \otimes \nabla_{b'} \sigma(x,x'))}{\sqrt{-g(x)}\sqrt{-g(x')}} \,.
\ee
The term $Q$ is the one introduced by Wald to preserve the
conservation of the stress-energy tensor. Wald \cite{Wald:1978pj}
calculated this term in the coincidence limit,
\be \label{Q}
Q=\frac{1}{12\pi^2}w_1(x,x) \,.
\ee
The term $C_{tt}$ handles the ambiguities in the definition of the
stress-energy tensor $T$ in curved spacetime.  We will adopt the
axiomatic definition given by Wald \cite{Wald:qft}, but there remains
the ambiguity of adding local curvature terms with arbitrary
coefficients.  From Ref.~\cite{Birrellbook} we find that these terms
include
\bml\label{eqn:12H}
\bea 
^{(1)}H_{ab} &=& 2R_{;ab} -2g_{ab}\Box R - g_{ab}R^2/2 + 2RR_{ab} \label{1H}\\
^{(2)}H_{ab} &=& R_{;ab} -\Box R_{ab} -g_{ab}\Box R/2
- g_{ab}R^{cd} R_{cd}/2  + 2R^{cd}R_{acbd} \label{2H}\,.
\elea
Thus in Eq.~(\ref{qinequality}) we must include a term given by a
linear combination of Eqs.~(\ref{1H}) and (\ref{2H}) to first order in $R$,
\be \label{localc}
C_{tt}=a\, ^{(1)}\!H_{tt}+b\, ^{(2)}\!H_{tt}=2aR_{,ii}-\frac{b}{2}(R_{tt,tt}+R_{ii,tt}-3R_{tt,ii}+R_{ii,jj}) \,,
\ee
where $a$ and $b$ are undetermined constants.\footnote{There are also
  ambiguities corresponding to adding multiples of the metric and the
  Einstein tensor to the stress tensor.  The first can be considered
  renormalization of the cosmological constant and the second
  renormalization of Newton's constant.  We will assume that these
  renormalization have been performed, and that the cosmological
  constant is considered part of the gravitational sector, so neither
  of these affects $T_{ab}$.}

From Ref.~\cite{Fewster:2007rh} we have the definition
\be \label{tilde}
\tilde{H}(x,x')=\frac{1}{2} \left[ H(x,x')+H(x',x)+iE(x,x') \right] \,,
\ee
where $iE$ is the antisymmetric part of the two-point function, which
we calculate in Sec.~\ref{sec:iE}.  We will use the
Fourier transform convention
\be \label{Fourier}
\hat{f}(k) \text{ or } f^{\wedge}[k]=\int_{-\infty}^\infty dxf(x) e^{ixk} \,.
\ee

We can now state the quantum inequality of Ref.~\cite{Fewster:2007rh},
\bea \label{qinequality}
\int_{-\infty}^\infty d\tau \, g(t)^2 \langle \Tren_{tt} \rangle_{\omega} (t,0) &\geq& -\int_0^\infty\frac{d\xi}{\pi} \left[  g \otimes g (\theta^* \Tsplit \tilde{H}_{(5)})(t,t') \right]^{\wedge} (-\xi,\xi) \nonumber\\
&&+\int_{-\infty}^\infty dt \, g^2(t) (-Q+C_{tt}) \,,
\eea
where the operator $\theta^*$ denotes the pullback of the function to
the geodesic,
\be
(\theta^*\Tsplit \tilde{H}_{(5)})(t,t') \equiv (\Tsplit \tilde{H}_{(5)})(\gamma(t),\gamma(t'))\,,
\ee
and the subscript $(5)$ means that we include
only terms through $j = 5$ in the sums of
Eq.~(\ref{hadamard}). However, as we proved in
Ref.~\cite{Kontou:2014eka}, terms of order $j >1$ make no contribution
to Eq.~(\ref{qinequality}).

Thus we can write  Eq.~(\ref{qinequality}) in our case as
\be\label{qinequality2}
\int_{-\infty}^\infty d\tau\,g(t)^2 \langle \Tren_{tt} \rangle (t,0)
\geq -B\,,
\ee 
where
\be\label{B}
B = \int_0^\infty\frac{d\xi}{\pi}\hat F(-\xi,\xi)
+\int_{-\infty}^\infty dt\,g^2(t) \left(Q-2aR_{,ii}-\frac{b}{2}(R_{tt,tt}+R_{ii,tt}-3R_{tt,ii}+R_{ii,jj})\right)\,,
\ee
\be\label{F}
F(t,t') = g(t)g(t')\Tsplit \tilde H_{(5)}((t,0),(t',0))\,,
\ee
and $\hat F$ denotes the Fourier transform in both arguments according to
Eq.~(\ref{Fourier}).

\section{Simplification of $\Tsplit$}
\label{sec:tsplit}

The $\Tsplit$ operator, Eq.~(\ref{tsplit}), can be written
\be \label{spti}
\Tsplit=\frac12\left[\partial_t \partial_{t'}+\sum_{i=1}^3\partial_i \partial_{i'}\right]\,.
\ee
To simplify it, we will define the following operator,
\be\label{barxd}
\nabla_{\bar{x}}^2=\nabla_x^2+2\sum_{i=1}^3\partial_i\partial_{i'}+\nabla_{x'}^2\,,
\ee
which in flat space would be the derivative with respect to the center point.
Then Eqs.~(\ref{spti}) and (\ref{barxd}) give
\bea
\Tsplit&=&\frac{1}{2}\left[\partial_t \partial_{t'}+\frac{1}{2}
  \left(\nabla_{\bar{x}}^2-\nabla_x^2-\nabla_{x'}^2\right)
\right]\nonumber\\
&=&\frac{1}{4}\left[\nabla_{\bar{x}}^2+\Box_x-\partial_t^2
+\Box_{x'}-\partial_{t'}^2+2\partial_t\partial_{t'}\right]\,,
\eea
where $\Box_x$ and $\Box_{x'}$ denote the D'Alembertian operator with
respect to $x$ and $x'$.  Because we are using Fermi coordinates and
are on the generating geodesic, the D'Alembertian and Laplacian
operators have the same form with respect to Fermi coordinates as they
do in flat space.  Then using
\be\label{bartd}
\partial_\tau^2=\frac14\left[\partial_t^2-2\partial_t \partial_{t'}+\partial_{t'}^2\right]\,,
\ee
we can write
\be
\Tsplit \tilde{H}=\frac{1}{4}\left[ \Box_x \tilde{H}
+ \Box_{x'} \tilde{H}+\nabla_{\bar{x}}^2 \tilde{H} \right]-\partial_\tau^2 \tilde{H}\,.
\ee
Consider the first term.  The function $H(x,x')$ obeys the equation of
motion in $x$ and so does $E(x,x')$.  Thus
\be \label{eqmd}
\Box_x \tilde{H}=\frac{1}{2}\Box_x H(x',x)\,.
\ee
The only asymmetrical part of $H$ comes from the $w_j$, so 
\be
H(x',x) = H(x,x') + \frac{1}{\pi^2}\sum_j(w_j(x',x)-w_j(x,x'))\sigma^j(x,x')\,.
\ee
and so we have
\be\label{EofMH1}
\Box_x  \tilde{H}
= \frac{1}{8\pi^2}\Box_x\sum_j(w_j(x',x)-w_j(x,x'))\sigma^j(x,x')\,.
\ee
Similarly,
\be\label{EofMH2}
\Box_{x'} \tilde{H}
= \frac{1}{8\pi^2}\Box_{x'}\sum_j(w_j(x,x')-w_j(x',x))\sigma^j(x,x')\,.
\ee
Adding together Eqs.~(\ref{EofMH1}) and (\ref{EofMH2}), we get
something which is symmetric in $x$ and $x'$ and vanishes in the
coincidence limit.  Following the analysis of \S3A of
Ref.~\cite{Kontou:2014eka}, such a term makes no contribution to
Eq.~(\ref{B}) and for our purposes we can take
\be \label{T}
\Tsplit \tilde{H}=\left[\frac{1}{4}\nabla_{\bar{x}}^2-\partial_\tau^2
\right] \tilde{H}\,.
\ee

\section{General computation of $E$}
\label{sec:iE}

The function $E$ is the advanced minus the retarded Green's function,
\be \label{theE}
E(x,x')=G_A(x,x')-G_R(x,x')\,,
\ee
and $iE$ is the imaginary, antisymmetric part of the two-point
function.  The Green's functions satisfy
\be \label{green}
\Box G(x,x')=\frac{\delta^{(4)}(x-x')}{\sqrt{-g}} \,.
\ee
Following Poisson, et al. \cite{Poisson:2011nh} and adjusting for different
sign and normalization conventions,
\be
G(x,x')=\frac{1}{4\pi} \left(2U(x,x')\delta(\sigma)+V(x,x')\Theta(-\sigma)\right)\,,
\ee
where $U(x,x') = \Delta^{1/2}(x,x')$ and $V(x,x')$ are smooth
biscalars.

For points $y$ null separated from $x'$, $V$ is called $\check V$
\cite{Poisson:2011nh} and satisfies
\be \label{vcheck}
\check V_{,a}\sigma^{,a}+\left[\frac{1}{2}\Box\sigma+2\right]\check V=-\Box U \,,
\ee
with all derivatives with respect to $y$.  Now $\check V$ is first
order in the curvature, so we will do the rest of the calculation as
though we were in flat space.  Under this approximation, we will
neglect coefficients which depend on the curvature, and also evaluate
curvature components at locations that would be relevant if we were in
flat space.  The distance between these locations and the proper
locations is first order in the curvature, so the overall inaccuracy
will always be second order in the curvature and its derivatives.

Thus we use $\sigma^{,a}=-2(y-x')^a$ and
$\Box\sigma=-8$ in Eq.~(\ref{vcheck}) to get
\be \label{vcheck2}
(y-x')^a \check V_{,a}(y)+\check V(y)=\frac{1}{2} \Box U(y) \,.
\ee
Now suppose we want to compute $\check V$ at some point $x''$.  We
need to integrate along the geodesic going from $x'$ to $x''$.  So let
$y=x'+\lambda(x''-x')$ and observe that
\be
\frac{d(\lambda \check V(y))}{d \lambda}
= \lambda\frac{ d \check V(y)}{d\lambda} + \check V(y)
= \lambda(x''-x')^a \check V_{,a} + \check V(y)
= (y-x')^a \check V_{,a} + \check V(y)
= \frac{1}{2} \Box U(y)\,,
\ee
so
\be\label{vcheckfinal}
\check V(x'',x') = \frac{1}{2} \int_0^1 d\lambda \Box U(y)\,.
\ee

The function $V$ obeys \cite{Poisson:2011nh}
\be
\Box V(x,x')=0 \,.
\ee
Consider points $x$ and $x'$ on the geodesic $\gamma$, which in the
flat-space approximation means they are separated only in time, and
let $\bar x = (x+x')/2$.  Then $V(x,x')$ can be found in terms of $V$
and its derivatives evaluated at the time $\bar t$ (the time component
of $\bar x$) using Kirchhoff's formula,
\be \label{Vu1}
V(x,x')=\frac{1}{4\pi} \int d\Omega \left[\check
  V(x'')+\frac{\tau}2\frac{\partial}{\partial r}\check
  V(x'')
+\frac{\tau}2\frac{\partial}{\partial t}\check V(x'') \right] \,,
\ee
where $\int d\Omega$ means to integrate over all unit vectors
$\hat\Omega$, and we now set
\be
x'' = \bar{x}+(\tau/2)\Omega
\ee
with the 4-vector $\Omega$ given by $\hat\Omega$ with unit time component.

Let us establish null-spherical coordinates $(u,v,\theta,\phi)$
with $u=t+r$, $v=t-r$, and the origin at $x'$.  Then $x''$ has
$u=\tau$, $v = 0$.  The derivative $\partial/\partial u$ can be
written $(\partial/\partial t + \partial/\partial r)/2$ and so
\be \label{Vu}
V(x,x')= \frac{1}{4\pi} \int d\Omega \, \frac{d}{du}\left[
u \check V(u\Omega/2,x')\right]_{u=\tau} \,.
\ee
From Eq.~(\ref{vcheckfinal}), 
\be
u \check V\big(\frac{u}2\Omega,x'\big)
= \frac{1}{2} \int_0^u du'(\Box U)(u'\Omega/2,x')
\ee
and so
\be\label{Vint}
V(x,x')=\frac{1}{8\pi} \int d\Omega\left[(\Box U)(\tau\Omega/2,x') \right] \,.
\ee

We are only interested in the first order of curvature, so we can
expand U, which is just the square root of the Van Vleck determinant,
to first order.  From Ref.~\cite{Visser:1992pz},
\be \label{deltaexp}
\Delta^{1/2}(x, x')=1-\frac{1}{2}
\int_0^1 ds (1-s)s R_{ab}(sx+(1-s)x')(x-x')^a (x-x')^b+O(R^2) \,,
\ee
so in the case at hand we can use
\be
U(x'') = \Delta^{1/2}(x'')=1-\frac{1}{2} \int_0^1 ds (1-s)s
R_{ab}(y) X^a X^b \,
\ee 
where $y=sx'' = (su'',sv'',\theta'', \phi'')$ is a point between 0 and
$x''$, and the tangent vector $X=dy/ds$.  We are interested in
$\Box_{x''} U(x'',0)$.  To bring the $\Box$ inside the integral, we
define $Y=sX = (su'',sv'',0,0)$, and
\be
\Box U(x'', 0) = -\frac{1}{2} \int_0^1 ds (1-s)s \Box_{x''}[R_{ab}(y) X^a X^b]
 = -\frac{1}{2} \int_0^1 ds (1-s)s \Box_{y}[R_{ab}(y) Y^a Y^b] \,.
\ee
For the rest of this section, all occurrences of $u$, $v$, $\theta$,
$\phi$, and derivatives with respect to these variables will refer to
these components of $y$ or $Y$.

Now we expand the D'Alembertian, in terms of angular derivatives,
derivatives in $u$, and derivatives in $v$,
\be
\Box=4\frac{\partial^2}{\partial v \partial u}-\frac{4}{u-v} \left(\frac{\partial}{\partial u}-\frac{\partial}{\partial v} \right)-\nabla^2_\Omega \,,
\ee
with 
\be
\nabla^2_\Omega = \frac{4}{(v-u)^2 \sin{\theta}}
\frac{\partial}{\partial \theta} \left(\sin{\theta}
\frac{\partial}{\partial \theta} \right)+\frac{4}{(v-u)^2
  \sin{\theta}^2}\frac{\partial^2}{\partial \phi^2} \,.
\ee
The angular integration in Eq.~(\ref{Vint}) annihilates the results of
$\nabla^2_\Omega$, so we have
\be \label{Vs}
V(x,x')=-\frac{1}{4\pi} \int d\Omega \int_0^1 ds s(1-s) \left[\partial_u \partial_v-\frac{1}{u-v} \left(\partial_u-\partial_v \right) \right](R_{ab}(y) Y^a Y^b) \,.
\ee
Outside the derivatives, we can take $v=0$ and change variables to $u=
s\tau$, giving
\bea
V(x,x')&=&-\frac{1}{4\pi\tau^3} \int d\Omega \int_0^\tau du (\tau-u)\left[
u\partial_u \partial_v- \partial_u+\partial_v \right](R_{ab}(y) Y^a Y^b)\\
&=&-\frac{1}{4\pi\tau^3} \int d\Omega \int_0^\tau du (\tau-u)
\partial_u[(u \partial_v- 1)(R_{ab}(y) Y^a Y^b)] \,.
\eea
We can integrate by parts with no surface contribution, giving
\bea
V(x,x')&=&\frac{1}{4\pi\tau^3} \int d\Omega \int_0^\tau du 
(1-u \partial_v)(R_{ab}(y) Y^a Y^b)\\
&=&\frac{1}{4\pi\tau^3} \int d\Omega \int_0^\tau du 
u^2 \left[ -u R_{uu,v}(y)-2 R_{uv}(y)+R_{uu}(y) \right]\nonumber \,.
\eea

Now
\be \label{RG}
R_{ab} = G_{ab}- (1/2)g_{ab}G \,,
\ee
where $G_{ab}$ is the Einstein tensor and $G$ its trace. Thus
\be\label{VG}
V(x,x')=\frac{1}{4\pi\tau^3} \int d\Omega \int_0^\tau du\,
u^2 \left[ -u G_{uu,v}(y)-2 G_{uv}(y)+(1/2)G(y)+G_{uu}(y) \right] \,.
\ee

Now define a vector field $Q_a(y) = G_{ab}(y)Y^b$.  Then
\be
Q_{a;c} =G_{ab;c}(y)Y^b+G_{ab}(y){Y^b}_{;c}\,.
\ee
We write the covariant derivative only because we are working in
null-spherical coordinates, rather than because of spacetime
curvature, which we are ignoring because we already have first order
quantities.

Since the covariant divergence of $G$ vanishes,
\be
g^{ac}Q_{a;c} =g^{ac} G_{ab}(y){Y^b}_{;c} \,.
\ee
In Cartesian coordinates, $Y^b = y^b$, and ${y^b}_{;c} = \delta^b_c$,
which means that (in any coordinate system).
\be
g^{ac}Q_{a;c} = G \,.
\ee
Explicit expansion gives
\be
g^{ac}Q_{a;c} = 2 (Q_{v,u} + Q_{u,v}) 
-\frac{4}{u-v} (Q_u-Q_v)
- \frac{4}{(v-u)^2}\left[ \frac1{\sin\theta}
\frac{\partial}{\partial \theta} (\sin{\theta} Q_\theta)
+\frac{1}{\sin\theta^2}Q_{\phi,\phi}\right] \,,
\ee
but the angular terms vanish on integration.  Now we expand
the derivatives in $u$ and $v$ and set $v=0$, giving
\blea
Q_{v,u}& = & uG_{uv,u}+ G_{uv}\\
Q_{u,v}& = & uG_{uu,v}+ G_{uv} \,,
\elea
so
\be\label{Qfinal}
\int d\Omega \,\left(2uG_{uv,u} + 2uG_{uu,v}+8G_{uv}- 4G_{uu}\right) = 
\int d\Omega \, G \,.
\ee
Substituting Eq.~(\ref{Qfinal}) into Eq.~(\ref{VG}), we find
\be
V(x,x')=\frac{1}{4\pi\tau^3} \int d\Omega \int_0^\tau du \,
u^2 \left[u G_{uv,u}(y)+2 G_{uv}(y)-G_{uu}(y) \right]
\ee
and integration by parts yields
\be
V(x,x')=\frac{1}{4\pi} \int d\Omega \left[G_{uv}(x'') -
\frac{1}{\tau^3} \int_0^\tau du \, u^2 \left(G_{uv}(y)+G_{uu}(y) \right)\right] \,.
\ee
Now
\bea
\int d\Omega \int_0^\tau du \, u^2 \left(G_{uv}(y)+G_{uu}(y) \right)
&=&\frac12\int d\Omega\int_0^\tau du \, u^2 \left(G_{tt}(y)+G_{tr}(y) \right)\nonumber\\
&=& \frac12\int d\Omega\int_0^\tau du \, u^2 \left(G^{tt}(y)-G^{tr}(y) \right)\,,
\eea
which is 4 times the total flux of $G^{ta}$ crossing inward through
the light cone.  Since this quantity is conserved, ${G^{ta}}_{;a}=0$,
we can integrate instead over a ball at constant time $\bar t$, giving
\be
4\int d\Omega\int_0^{\tau/2} dr\, r^2 G^{tt}(\bar x + r\Omega )
= \frac{\tau^3}2\int d\Omega\int_0^1 ds\, s^2 G^{tt}(\bar x + s(\tau/2)\Omega)
\ee
so
\be
V(x,x')=\frac{1}{8\pi} \int d\Omega \left[\frac{1}{2} \left[ G_{tt}(x'')-G_{rr}(x'') \right]
-\int_0^1 ds\, s^2 G_{tt}(x''_s)\right] \,,
\ee
where $x''_s=\bar{x}+s(\tau/2)\Omega$, and
\bea
G_R(x,x')=\Delta^{1/2}(x,x') \frac{\delta(\sigma)}{2\pi}+\frac{1}{32 \pi^2} \int d\Omega\bigg\{\frac{1}{2} \left[ G_{tt}(x'')-G_{rr}(x'') \right]-\int_0^1 ds \, s^2 G_{tt}(x''_s) \bigg\} \,,
\eea
\bea \label{finaliE}
E(x,x')=\Delta^{1/2}(x,x') \frac{\delta(\tau-|\bx-\bx'|)-\delta(\tau+|\bx-\bx'|)}{4\pi|\bx-\bx'|}&+&\frac{1}{32 \pi^2} \int d\Omega\bigg\{ \frac{1}{2} \left[ G_{tt}(x'')-G_{rr}(x'') \right]\nonumber\\
&&-\int_0^1 ds \,  s^2  G_{tt}(x''_s) \bigg\} \sgn{\tau} \,.
\eea

\section{Computation of $\tilde{H}$}
\label{sec:H}

We now need to compute $\tilde H(x,x')$ and apply $\Tsplit$.  First we
consider the term in $\tilde H(x,x')$ that has no dependence on the
curvature.  It has the same form as it would in flat space
\cite{Fewster:2007rh,Kontou:2014eka},
\be\label{H-1}
\tilde H_{-1}(x,x')=H_{-1}(x,x')=\frac{1}{4\pi^2\sigma_+(x,x')}\,.
\ee
In Sec.~\ref{sec:tsplitH}, we will apply the fully general $\Tsplit$
from Eq.~(\ref{T}) with $\nabla_{\bar{x}}$ defined in
Eq.~(\ref{barxd}) to $\tilde H_{-1}(x,x')$.

All the remaining terms that we need are first order in the curvature,
so for these it is sufficient to take $\nabla_{\bar{x}}$ as the
flat-space Laplacian with respect to the center point, $\bar{x}$.
For this we only need to compute $\tilde H$ at positions given by time
coordinates $t$ and $t'$ but the same spatial position.

As we discussed, we only need to keep terms in $\tilde H$ with powers
of $\tau$ up to $\tau^2$, but we need $E$ exactly.  The terms from $H$
alone give a function whose Fourier transform does not decline fast
enough for positive $\xi$ for the integral in Eq.~(\ref{B}) to
converge.  Thus we extract the leading order terms from $iE$ and
combine these with the terms from $H$.  This combination gives a
result that has the appropriate behavior after the Fourier transform.

Following the notation of Ref~\cite{Kontou:2014eka}, we let
$H_j(t,t')$, $j=0,1,\ldots$, denote the term in $H$ involving
$\tau^{2j}$ (with or without $\ln\tau$), and $H_{(j)}$ denote the sum
of all terms from $H_{-1}$ through $H_j$.  We will split up $E(x,x')$
in similar fashion, define a ``remainder term''
\be
R_j = E - \sum_{k=-1}^j E_k\,,
\ee
and let
\blea
\tilde
H_{j}(x,x') &=& \frac12\left[H_j(x,x') + H_j(x',x) + iE_j(x,x')\right]\\
\tilde H_{(j)}(x,x') &=& \frac12\left[H_{(j)}(x,x') + H_{(j)}(x',x) + iE(x,x')\right]\,.
\elea

\subsection{Terms with no powers of $\tau$}

First we want to calculate the zeroth order of the Hadamard
series. The Hadamard coefficients are given by the Hadamard recursion
relations, which are the solutions to
\be
\Box H(x,x')=0, w_0=0 \,.
\ee
The recursion relations for the massless field in a curved background are \cite{Fewster:2007rh}
\be \label{recursion1}
\Box \Delta^{1/2}+2 v_{0,a} \sigma^{,a}+4 v_0+v_0 \Box \sigma=0 \,,
\ee
\be \label{recursion2}
\Box v_j +2(j+1) v_{j+1,a}\sigma^{,a}-4j(j+1)v_{j+1}+(j+1)v_{j+1}\Box \sigma=0 \,.
\ee
To find the zeroth order of the Hadamard series we need only
$v_0(x,x')$, which we find by integrating Eq.~(\ref{recursion1}) along
the geodesic from $x'$ to $x$.  Since we are computing a first-order
quantity, we can work in flat space by letting $y'=x'+\lambda(x-x')$
and using the first-order formulas $\Box \sigma=-8$ and
$\sigma^{,a}=-2 (y'-x')^a$.  From Eq.~(\ref{recursion1}), we have
\be
(y'-x')^a v_{0,a}+v_0=\frac{1}{4} \Box \Delta^{1/2} (y',x') \,,
\ee
and thus
\be\label{v01}
v_0(x,x')=\frac{1}{4} \int_0^1 d\lambda (\Box \Delta^{1/2}) (x'+\lambda(x-x'),x') \,.
\ee
by the same analysis as Eq.~(\ref{vcheckfinal}).

Using the expansion for $\Delta^{1/2}$ from Eq.~(\ref{deltaexp}) gives
\bea
v_0(x,x')&=&-\frac{1}{8}\int_0^1 d\lambda \int_0^1 ds (1-s)s
\Box_{y'}[R_ {ab}(sy'+ (1 -s)x') (y' -x') ^a (y' -x') ^b]\nonumber\\
&=&-\frac{1}{8} \int_0^1 d\lambda \int_0^1 ds(1-s)s \bigg[ (\lambda s)^2 (\Box R_{ab})(x'+s\lambda(x-x'))(x-x')^a (x-x')^b\nonumber\\
&&+2\lambda s R_{,b} (x'+s\lambda(x-x')) (x-x')^b+2 R(x'+s\lambda (x-x')) \bigg] \,.
\eea
We can combine the $s$ and $\lambda$ integrals by defining a new variable $\sigma=s\lambda$
\bea \label{lambdasigma}
&&\int_0^1 d\lambda \int_0^1 ds (1-s)s f(\lambda s)=\int_0^1 d\lambda \int_0^\lambda d\sigma \left(\frac{\sigma}{\lambda^2}-\frac{\sigma^2}{\lambda^3} \right) f(\sigma) \\
&&=\int_0^1 d\sigma \, f(\sigma) \int_\sigma^1 d\lambda \left(\frac{\sigma}{\lambda^2}-\frac{\sigma^2}{\lambda^3} \right)=\int_0^1 d\sigma \, f(\sigma) \left[ -\frac{\sigma}{\lambda}+\frac{\sigma^2}{2\lambda^2} \right]_\sigma^1=\frac{1}{2}\int_0^1 d\sigma \, f(\sigma) (1-\sigma)^2 \nonumber \,.
\eea
Then, changing $\sigma$ to $s$, we find
\bea\label{v0s}
v_0(x,x')&=&-\frac{1}{16}\int_0^1 ds (1-s)^2  \bigg[s^2 (\Box R_{ab})(x'+s(x-x'))(x-x')^a (x-x')^b\nonumber\\
&&+2  s R_{,b} (x'+s(x-x')) (x-x')^b+2 R(x'+s (x-x')) \bigg] \,.
\eea
or when the two points are on the geodesic,
\be
v_0(t,t')=-\frac{1}{16} \int_0^1 ds (1-s)^2 \bigg[ s^2 (\Box R_{tt})(x'+s \tau) \tau^2+4s \eta^{cd}  R_{ct,d}(x'+s \tau)\tau+2 R(x'+s \tau) \bigg] \,.
\ee
In the second term we use the contracted Bianchi identity, $\eta^{cd}
R_{ct,d}= R_{,t}/2$, giving
\bea
&&2\int_0^1 ds (1-s)^2 s\tau R_{,t}(x'+s\tau)= 2\int_0^1 ds (1-s)^2 s \frac{d}{ds} R(x'+s\tau) \nonumber\\
&&=-2\int_0^1 ds (1-s)(1-3s)R(x'+s\tau)\,,
\eea
so the final expression for $v_0$ is
\be \label{v0}
v_0(t,t')=-\frac{1}{16} \int_0^1 ds  (1-s) \bigg[ s^2 (1-s) \Box R_{tt} (\bar{x}+(s-1/2)\tau) \tau^2+4 s R(\bar{x}+(s-1/2)\tau) \bigg] \,.
\ee 
To calculate $H_0$ we only need the zeroth order in $\tau$ from
$v_0$, so the first term does not contribute.  In the second term, we
make a Taylor series expansion,
\be\label{Rt1}
R(\bar{x}+(s-1/2)\tau)=R(\bar{x})+R_{,t}(\bar{x})\tau (s-1/2)\tau+\frac{1}{2}R_{,tt}(\bar{x})\tau^2 (s-1/2)^2+O(\tau^3) \,,
\ee
but only the first term is relevant here.  Thus
\be\label{v0final}
v_0(t,t')=-\frac{1}{4}\int_0^1 ds (1-s)s R(\bar{x})=-\frac{1}{24} R(\bar{x}) \,.
\ee
We also need to expand the Van Vleck determinant appearing in the the
Hadamard series.  From Eq.~(\ref{deltaexp}),
\be \label{vanvleck}
\Delta^{1/2}(t,t')=1-\frac{1}{12}R_{tt}(\bar{x})\tau^2-\frac{1}{480}R_{tt,tt}(\bar{x})\tau^4+O(\tau^6)\,.
\ee
Keeping the first order term from Eq.~(\ref{vanvleck}) and using
Eq.~(\ref{v0final}), we have
\be \label{had0}
H_0(x,x')=\frac{1}{48\pi^2} \left[R_{tt}(\bar{x})-\frac{1}{2}R(\bar{x}) \ln{(-\tau_-^2)} \right] \,.
\ee
Now we can add the $H_0(x',x)$ which is the same except that $t$ and $t'$ interchange
\be \label{H0}
H_0(x,x')+H_0(x',x)=\frac{1}{24\pi^2} \left[R_{tt}(\bar{x})-R(\bar{x}) \ln{|\tau|} \right] \,.
\ee

Next we must include $E$ from Eq.~(\ref{finaliE}).  We can expand the
components of the Einstein tensor around $\bar{x}$,
\be
G_{ab}(x'')=G_{ab}(\bar{x})+G_{ab}^{(1)}(x'') \,,
\ee
where $G_{ab}^{(1)}$ is the remainder of the Taylor series
\be \label{rem1}
G_{ab}^{(1)}(x'')=G_{ab}(x'')-G_{ab}(\bar{x})=\int_0^{\tau/2} dr \,G_{ab,i}(\bar{x}+r\Omega)\Omega^i \,.
\ee
Then from Eq.~(\ref{finaliE}) and using $\int d\Omega\, \Omega^i=0$ and
$\int d\Omega\, \Omega^i \Omega^j=(4\pi/3) \delta^{ij}$ we have
\bea \label{E0}
E_0(x,x')&=&\frac{1}{8 \pi} \left\{ \frac{1}{2}G_{tt}(\bar{x})-\frac{1}{6}G_{ii}(\bar{x})-\int_0^1 ds \, s^2 G_{tt}(\bar{x}) \right\} \sgn{\tau}\nonumber\\
&=& \frac{1}{48 \pi}  G(\bar{x})\sgn{\tau} =-\frac{1}{48 \pi} R(\bar{x})\sgn{\tau}
\eea
and
\bea \label{R0}
R_0(x,x')=\frac{1}{32\pi^2}\int d\Omega \left\{ \frac{1}{2}\left[G_{tt}^{(1)}(x'')-G_{rr}^{(1)}(x'')\right]-\int_0^1 ds\,s^2 G_{tt}^{(1)}(x''_s) \right\} \sgn{\tau} \,.
\eea
Using
\be \label{log}
2\ln{|\tau|}+\pi i\sgn{\tau}=\ln{(-\tau_-^2)} \,,
\ee
we combine Eqs.~(\ref{H0}) and (\ref{E0}) to find
\be \label{Ht0}
\tilde{H}_0(t,t')=\frac{1}{48\pi^2} \left[ R_{tt}(\bar{x})-\frac{1}{2}R(\bar{x}) \ln{(-\tau_-^2)} \right] \,.
\ee
Combining all terms through order 0 gives
\be \label{Ht-10}
\tilde{H}_{(0)}(t,t')=\tilde{H}_{-1}(t,t')+\tilde{H}_0(t,t')+\frac{1}{2}i R_0(t,t') \,.
\ee

\subsection{Terms of order $\tau^2$}

Now we compute the terms of order $\tau^2$ in $H$ and $E$. To find
$v_0$ at this order we take Eqs.~(\ref{v0}) and (\ref{Rt1}) and
include terms through second order in $\tau$,
\bea \label{v0ab}
v_0(x,x')&=&-\frac{1}{24}R(\bar{x})-\frac{1}{16} \int_0^1 ds (1-s) \left[ s^2(1-s) \Box R_{tt}(\bar{x})+2s(s-1/2)^2 R_{,tt}(\bar{x})\right]\tau^2+ \dots \nonumber\\
&=&-\frac{1}{24}R(\bar{x})-\frac{1}{480} \left(\Box R_{tt}(\bar{x})+\frac{1}{2}R_{,tt}(\bar{x}) \right)\tau^2+\dots \,.
\eea

Next we need $v_1$ but since it is multiplied by $\tau^2$ in $H$ we
need only the $\tau$ independent term. From Eq.~(\ref{recursion2})
\be
\Box v_0+2v_{1,a} \sigma^{,a}+v_1 \Box \sigma=0 \,,
\ee
At $x=x'$, $ \sigma^{,a}=0$ so
\be\label{v1lim}
v_1(x,x)=\frac{1}{8} \lim_{x \to x'} \Box_x v_0(x,x') \,.
\ee
Using Eq.~(\ref{v0s}) in  Eq.~(\ref{v1lim}), the only terms that
survive in the coincidence limit are those that have no powers of
$x-x'$ after differentiation, so
\be \label{v1}
v_1(x,x) = -\frac{1}{16} \int_0^1 ds (1-s)^2 s^2 \Box R(\bar{x})=-\frac{1}{480} \Box R(\bar{x}) \,.
\ee
Equations~(\ref{v0}), (\ref{v0ab}) and (\ref{v1}) agree with
Ref.~\cite{Decanini:2005gt} if we note that their expansions are
around $x$ instead of $\bar{x}$.

The $w_1$ at coincidence is given by Ref.~\cite{Wald:1978pj},
\be \label{w1}
w_1(x,x)=-\frac{3}{2} v_1(x,x)=\frac{1}{320} \Box R(\bar{x}) \,.
\ee

Combining Eqs.~(\ref{v0ab}), (\ref{v1}), and (\ref{w1}), and the
fourth order term from the Van Vleck determinant of
Eq.~(\ref{vanvleck}), and keeping in mind that $\sigma=-\tau^2$ when
both points are on the geodesic, we find
\be
H_1(x,x')=\frac{1}{640\pi^2} \bigg[\frac{1}{3}R_{tt,tt}(\bar{x})-\frac{1}{2} \Box R(\bar{x})
-\frac{1}{3}\left(\Box R_{ii}(\bar{x})+\frac{1}{2}R_{,tt}(\bar{x})\right)\ln{(-\tau_-^2)} \bigg]\tau^2  \,.
\ee
Then $H_1(x',x)$ is given by symmetry so
\be \label{H1}
H_1(x,x')+H_1(x',x)=\frac{1}{160\pi^2} \left[\frac{1}{6}R_{tt,tt}(\bar{x})-\frac{1}{4}\Box R(\bar{x})-\frac{1}{3}\left(\Box R_{ii}(\bar{x})+\frac{1}{2}R_{,tt}(\bar{x})\right)\ln{|\tau|}  \right]\tau^2 \,.
\ee

The calculation of $E_1$ is similar to $E_0$, but now we have to include more terms to the Taylor expansion,
\be
G_{ab}(x'')=G_{ab}(\bar{x})+\frac{\tau}{2}G_{ab,i}(\bar{x})\Omega^i+\frac{\tau^2}{8}G_{ab,ij}\Omega^i \Omega^j(\bar{x})+G_{ab}^{(3)}(x'') \,,
\ee
where the remainder of the Taylor series is
\be \label{rem3}
G_{ab}^{(3)}(x'')=\frac{1}{2}\int_0^{\tau/2} dr G_{ab,ijk}(\bar{x}+r\Omega) \left( \frac{\tau}{2}-r\right)^2 \Omega^i \Omega^j \Omega^k \,.
\ee
Then from Eq.~(\ref{finaliE}) and using that $\int d\Omega \, \Omega^i=\int d\Omega \, \Omega^i \Omega^j \Omega^k=0$, $\int d\Omega \, \Omega^i \Omega^j=4\pi/3 \delta^{ij}$ and $\int d\Omega \, \Omega^i \Omega^j \Omega^k \Omega^l=(4\pi/15) (\delta^{ij} \delta^{kl}+\delta^{ik}\delta^{jl} +\delta^{il}\delta^{jk})$ we have
\bea 
E_1(x,x')&=&-\frac{1}{192 \pi} \left[ \frac{1}{10} G_{ii,jj}(\bar{x})+\frac{1}{5} G_{ij,ij}(\bar{x}) -\frac{1}{2} G_{tt,ii}(\bar{x})+\int_0^1 ds \, s^4 G_{tt,ii}(\bar{x}) \right]\tau^2 \sgn{\tau} \nonumber\\
&=&-\frac{1}{320\pi} \left[ \frac{1}{6} G_{ii,jj}(\bar{x})+\frac{1}{3} G_{ij,ij}(\bar{x})-\frac{1}{2} G_{tt,ii} (\bar{x}) \right] \tau^2 \sgn{\tau} \,.
\eea
Using the conservation of the Einstein tensor, $0 = \eta^{ab} G_{ia,b} = G_{it,t} - G_{ij,j}$ and 
$0 = \eta^{ab} G_{ta,b} = G_{tt,t} - G_{it,i}$ we can write
\be
G_{ij,ij}(\bar{x})=G_{tt,tt}(\bar{x}) \,.
\ee
So
\bea \label{E1}
E_1(x,x')&=&-\frac{1}{960\pi}  \left(\frac{1}{2}G_{ii,jj}(\bar{x})+ G_{tt,tt}(\bar{x})-\frac{3}{2} G_{tt,ii} (\bar{x})\right) \tau^2 \sgn{\tau}\nonumber\\
&=&-\frac{1}{960\pi}  \left( \Box R_{ii}(\bar{x})+\frac{1}{2} R_{,tt}(\bar{x}) \right) \tau^2 \sgn{\tau}
\eea
and
\be \label{R1}
R_1(x,x')=\frac{1}{32\pi^2} \int d\Omega \bigg\{ \frac{1}{2} \left[G_{tt}^{(3)}(x'')-G_{rr}^{(3)}(x'') \right]-\int_0^1 ds \, s^2 G_{tt}^{(3)}(x''_s)  \bigg\} \sgn{\tau} \,.
\ee
To calculate $\tilde{H}_1$, we combine Eqs.~(\ref{H1}) and (\ref{E1})
and use Eq.~(\ref{log}) to get
\be \label{Ht1}
\tilde{H}_1(x,x')=\frac{\tau^2}{640\pi^2} \left[ \frac{1}{3}R_{tt,tt}(\bar{x})-\frac{1}{2} \Box R(\bar{x})-\frac{1}{3} \left( \Box R_{ii}(\bar{x})+\frac{1}{2} R_{,tt}(\bar{x}) \right) \ln{(-\tau_-^2)} \right] \,.
\ee
All terms through order 1 are then given by
\be \label{Ht-11}
\tilde{H}_{(1)}(t,t')=\tilde{H}_{-1}(t,t')+\tilde{H}_0(t,t')+\tilde{H}_1(t,t')+\frac{1}{2}iR_1(t,t') \,.
\ee

\section{The $\Tsplit \tilde{H}$}
\label{sec:tsplitH}

We can easily take the derivatives of $\tilde{H}_0$ and $\tilde{H}_1$
using Eq.~(\ref{T}), because they are already first order in
$R$. However in the case of the term $\nabla_{\bar{x}}^2
\tilde{H}_{-1}$ we have to proceed more carefully. From
Eqs.~(\ref{barxd}) and (\ref{H-1}) we have
\bea \label{TH-1}
\nabla_{\bar{x}}^2 \tilde{H}_{-1}&=&\frac{1}{4\pi^2} \sum_{i=1}^3 \left( \frac{\partial^2}{\partial (x^i)^2}+2\frac{\partial}{\partial x^i} \frac{\partial}{\partial x'^i}+\frac{\partial^2}{\partial (x'^i)^2} \right)\left(\frac{1}{\sigma_+}\right)\nonumber\\
&=&-\frac{1}{4\pi^2\sigma_+^2}  \sum_{i=1}^3 \left( \frac{\partial^2\sigma}{\partial (x^i)^2}+2\frac{\partial^2 \sigma}{\partial x^i\partial x'^i}+\frac{\partial^2 \sigma}{\partial (x'^i)^2} \right) \,,
\eea
where we used $\partial \sigma/ \partial x^i=\partial \sigma/
\partial x'^i=0$ when the two points are on the geodesic. From \cite{Decanini:2005gt}, after we shift the Taylor series so that the Riemann tensor is evaluated at $\bar{x}$, we have 
\bml\label{d2sigma}\bea
\frac{\partial^2 \sigma}{\partial (x^i)^2}&=&=-2\eta_{ii}-\frac{2}{3}R_{itit}(\bar{x})\tau^2-\frac{1}{2} R_{itit,t}(\bar{x})\tau^3-\frac{1}{5} R_{itit,tt}\tau^4+O(\tau^5) \label{sigma1}\\
\frac{\partial^2 \sigma}{\partial (x'^i)^2}&=&=-2\eta_{ii}-\frac{2}{3}R_{itit}(\bar{x})\tau^2+\frac{1}{2} R_{itit,t}(\bar{x})\tau^3-\frac{1}{5} R_{itit,tt}\tau^4+O(\tau^5) \label{sigma2}\\
\frac{\partial^2 \sigma}{\partial x^i\partial x'^i} &=&2\eta_{ii}-\frac{1}{3}R_{itit}(\bar{x})\tau^2-\frac{7}{40} R_{itit,tt}\tau^4+O(\tau^5)\label{sigma3} \,.
\elea
From Eqs.~(\ref{TH-1}) and (\ref{d2sigma}), and using $R_{itit}=-R_{tt}$ we have
\be \label{TH}
\nabla_{\bar{x}}^2 \tilde{H}_{-1}=-\frac{1}{4\pi^2} \left[\frac{2}{\tau^2}R_{tt}(\bar{x})+\frac{3}{4} R_{tt,tt}(\bar{x}) \right] \,.
\ee
From Eqs.~(\ref{F}) and (\ref{T}), we need to compute
\be
\int_0^\infty \frac{d\xi}{\pi}\hat F(-\xi,\xi')\,,
\ee
where
\be
F(t,t') = g(t) g(t')\left[\frac{1}{4}\nabla_{\bar{x}}^2 \tilde H_{(0)}(t,t')
-\partial_\tau^2  \tilde{H}_{(1)}(t,t') \right]\,.
\ee
Using
Eqs.~(\ref{H-1}), (\ref{R0}), (\ref{Ht0}), (\ref{Ht-10}), (\ref{R1}), (\ref{Ht1}), (\ref{Ht-11}) and (\ref{TH})
we can combine all terms in $F$ to write
\be
F(t,t') = g(t) g(t')\sum_{i=1}^6 f_i(t,t')\,,
\ee
with
\blea
f_1&=&\frac{3}{2\pi^2\tau_-^4}\\\
f_2&=&\frac{1}{48\pi^2\tau_-^2}[R_{ii}(\bar{x})-7 R_{tt}(\bar{x})]\\
f_3&=&\frac{1}{384\pi^2} \left[ \frac{1}{5} R_{tt,tt}(\bar{x})+\frac{1}{5}R_{ii,tt}(\bar{x})-R_{tt,ii}(\bar{x})+\frac{3}{5}R_{ii,jj}(\bar{x}) \right] \ln{(-\tau_-^2)}\\
f_4&=&\frac{1}{320\pi^2}\bigg[-\frac{43}{3}R_{tt,tt}(\bar{x})+\frac{7}{6} R_{tt,ii}(\bar{x})-\frac{1}{2}R_{ii,jj}(\bar{x}) \bigg] \\
f_5&=&\frac{1}{256\pi^2} \int d\Omega\,\nabla_{\bar{x}}^2 \left\{ \frac{1}{2} \left[ G_{tt}^{(1)}(x'')-G_{rr}^{(1)}(x'') \right] -\int_0^1 dss^2 G_{tt}^{(1)}(x''_s) \right\} i\sgn{\tau} \label{f5}\\
f_6&=& -\frac{1}{64\pi^2} \int d\Omega\,\partial_{\tau}^2  \bigg\{ \frac{1}{2} \left[G_{tt}^{(3)}(x'')-G_{rr}^{(3)}(x'') \right]-\int_0^1 ds s^2 G_{tt}^{(3)}(x''_s)  \bigg\} i\sgn{\tau} \label{f6} \,.
\elea

\section{The quantum inequality}
\label{sec:QI}

We want to calculate the quantum inequality bound $B$, given
by Eq.~(\ref{B}).  We can write it
\be
B=\sum_{i=1}^8 B_i\,,
\ee
where
\blea
B_i&=&\int_0^\infty  \frac{d\xi}{\pi} \int_{-\infty}^\infty dt
\int_{-\infty}^\infty dt' g(t) g(t') f_i(t,t') e^{i\xi(t'-t)}\nonumber\\
&=&\int_0^\infty \frac{d\xi}{\pi} \int_{-\infty}^\infty d\tau
\int_{-\infty}^\infty d\bar{t}\,
g(\bar{t}-\frac{\tau}{2})g(\bar{t}+\frac{\tau}{2})f_i(\bar t,\tau)e^{-i\xi \tau}
\qquad i = 1\ldots6\\
\label{B7}B_7 &=& \int_{-\infty}^\infty dt\,g^2(t) Q(t)
= \frac{1}{3840\pi^2}\int_{-\infty}^\infty dt\,g^2(t)\Box R(\bar{t})\\
\label{B8}
B_8 &=& -\int_{-\infty}^\infty dt\,g^2(t)\left[2aR_{,ii}(\bar{x})-\frac{b}{2}(R_{tt,tt}(\bar{x})+R_{ii,tt}(\bar{x})-3R_{tt,ii}(\bar{x})+R_{ii,jj}(\bar{x}))\right] 
\elea
using Eqs.~(\ref{Q}), (\ref{localc}), (\ref{B}) and (\ref{w1}).  The first 6 terms
have exactly the same $\tau$ dependence as the corresponding terms in
Ref.~\cite{Kontou:2014eka}. So the Fourier transform proceeds in the
same way, except that instead of dependence on the potential and its
derivatives, we have dependence on the Ricci tensor and its
derivatives.  After the Fourier transform, we see that $B_4$ and $B_7$
have exactly the same form so we merge them in one term. Thus
\be\label{BforR}
B = \frac{1}{16\pi^2}\left[ I_1
+\frac{1}{12} I_2
-\frac{1}{12} I_3
+\frac{1}{240} I_4
+\frac{1}{16\pi} I_5
-\frac{1}{4\pi} I_6\right]
-I_7\,,
\ee
where
\bml\label{I}\bea
I_1&=&\int_{-\infty}^{\infty} dt\,g''(t)^2 \\
I_2&=&
\int_{-\infty}^\infty d\bar{t}\,[R_{ii}(\bar{x})-7R_{tt}(\bar{x})]
(g(\bar{t}) g''(\bar{t}) - g'(\bar{t}) g'(\bar{t}))\\
\label{I3}I_3  &=&\int_{-\infty}^{\infty} d\tau\,\ln{|\tau|}\sgn{\tau}
\int_{-\infty}^{\infty} d\bar t\,\bigg[\frac{1}{5} R_{tt,tt}(\bar{x})+\frac{1}{5}R_{ii,tt}(\bar{x})-R_{tt,ii}(\bar{x})  \nonumber \\
&&+\frac{3}{5}R_{ii,jj}(\bar{x})\bigg] g(\bar{t}-\frac{\tau}{2})g'(\bar{t}+\frac{\tau}{2})\\
\label{I4}I_4 &=& \int_{-\infty}^{\infty}d\bar{t}\,
g(\bar{t})^2\bigg[-171 R_{tt,tt}(\bar{x})-R_{ii,tt}(\bar{x})+13R_{tt,ii}(\bar{x})-5R_{ii,jj}(\bar{x})\bigg]
\\
\label{I5}I_5 &=& \int_{-\infty}^\infty d\tau \, \frac{1}{\tau}
\int_{-\infty}^\infty d\bar{t}\,g(\bar{t}-\tau/2)g(\bar{t}+\tau/2) \int d\Omega\,\nabla_{\bar{x}}^2 \bigg\{ \frac{1}{2}\left[G_{tt}^{(1)}(x'')-G_{rr}^{(1)}(x'')\right] \nonumber\\
&&-\int_0^1 ds \, s^2 \left[G_{tt}^{(1)}(x''_s)\right] \bigg\} \sgn{\tau}
\\
\label{I6}I_6&=&\int_{-\infty}^\infty d\tau
\int_{-\infty}^\infty d\bar t\,
\partial_\tau^2 \left[\frac{1}{\tau} g(\bar{t}-\tau/2)g(\bar{t}+\tau/2)  \right]
\int d\Omega\,\bigg\{ \frac{1}{2} \left[G_{tt}^{(3)}(x'')-G_{rr}^{(3)}(x'') \right]\nonumber\\
&&-\int_0^1 ds \, s^2 G_{tt}^{(3)}(x''_s) \bigg\}\sgn{\tau}\\
\label{I7}I_7 &=&\int_{-\infty}^\infty dt\,g^2(t)\left[2aR_{,ii}(\bar{x})-\frac{b}{2}(R_{tt,tt}(\bar{x})+R_{ii,tt}(\bar{x})-3R_{tt,ii}(\bar{x})+R_{ii,jj}(\bar{x}))\right]\,.
\elea
If we only know that the Ricci tensor and its derivatives are bounded,
as in Eq.~(\ref{Rmax}), we can restrict the magnitude of each term of
Eq.~(\ref{BforR}). We start with the second term
\bea
|I_2| &\leq& \int_{-\infty}^\infty d\bar{t}\, \left|R_{ii}(\bar{x})-7R_{tt}(\bar{x})\right|
|g(\bar{t}) g''(\bar{t}) - g'(\bar{t}) g'(\bar{t})|  \nonumber\\
&\leq& 10 \Rmax \int_{-\infty}^\infty d\bar{t} [g(\bar{t}) |g''(\bar{t})| + g'(\bar{t})^2] \,.
\eea
Terms $I_3$, $I_4$ and $I_7$ are similar. Since Eq.~(\ref{Rmax}) holds
regardless of rotation, we can write $G_{rr}$ in terms of radial and
as azimuthal components of $R$ to find $|G_{rr}| < 2\Rmax$, and
similarly $|G_{tt}|<2\Rmax$. Using these results and Eq.~(\ref{rem1})
for the remainder we have
\bea
&& \left| \int d\Omega\,\nabla_{\bar{x}}^2 \bigg\{ \frac{1}{2}\left[G_{rr}^{(1)}(x'')-G_{tt}^{(1)}(x'')\right]+\int_0^1 ds \, s^2 G_{tt}^{(1)}(x''_s) \bigg\} \right|  \nonumber\\
&& \leq \frac{|\tau|}{2} \int d\Omega \left\{ \frac{1}{2} \left[
  |\nabla^2 G_{rr,i}(\bar x)|+|\nabla^2 G_{tt,i}(\bar x)| \right]+\int_0^1 ds \, s^3 |\nabla^2 G_{tt,i}(\bar x)| \right\} |\Omega^i| \nonumber\\
&& \leq \Rmax''' \frac{15|\tau|}{4} \sum_i \int d\Omega |\Omega^i|=\frac{45\pi}{2} |\tau| \Rmax''' \,.
\eea
For $I_6$ we use Eq.~(\ref{rem3}) for the remainder
\bea
&&\left| \int d\Omega\,\bigg\{ \frac{1}{2} \left[G_{rr}^{(3)}(x'')-G_{tt}^{(3)}(x'') \right]+\int_0^1 ds \, s^2 G_{tt}^{(3)}(x''_s) \bigg\} \right|  \nonumber\\
&& \leq \frac{|\tau|^3}{48} \int d\Omega \left\{ \frac{1}{2} \left[ |G_{rr,ijk}(\bar x)|+|G_{tt,ijk}(\bar x)| \right]+\int_0^1 ds \,s^5 |G_{tt,ijk}(\bar x)| \right\} |\Omega^i||\Omega^j||\Omega^k| \nonumber\\
&& \leq \Rmax''' \frac{7 |\tau|^3}{144} \sum_{i,j,k}\int d\Omega |\Omega^i| |\Omega^j| |\Omega^k|=\frac{7( 2 \pi+1)}{24} |\tau|^3 \Rmax''' \,.
\eea
After we bound all the terms and calculate the derivatives in $I_6$ we can define
\bml\label{J17}\bea
J_2&=&\int_{-\infty}^\infty dt\left[g(t)|g''(t)|+g'(t)^2\right]\\
J_3&=&\int_{-\infty}^\infty dt \int_{-\infty}^\infty dt' |g'(t')|g(t) |\!\ln{|t'-t|}| \\ J_4&=&\int_{-\infty}^\infty dt\,g(t)^2\\
J_5&=&\int_{-\infty}^\infty dt \int_{-\infty}^\infty dt' g(t)g(t') \\
J_6&=&\int_{-\infty}^\infty dt \int_{-\infty}^\infty dt' |g'(t')|g(t)|t'-t|\\
J_7&=&\int_{-\infty}^\infty dt \int_{-\infty}^\infty dt' 
\left [g(t)|g''(t')| +g'(t)g'(t')\right] (t'-t)^2
\elea
and find
\blea
|I_2| &\le & 10 \Rmax J_2\\
|I_3| &\le &\frac{46}{5} \Rmax'' J_3\\
|I_4| &\le & 258 \Rmax''J_4\\
|I_5|&\le & \frac{45\pi}{2} \Rmax''' J_5\\
|I_6|&\le &\frac{7(2\pi+1)}{48}\Rmax'''\left(4J_5+4J_6+J_7\right)\\
|I_7|&\le & (24|a|+11|b|)\Rmax''J_4\,.
\elea
Thus the final form of the inequality is
\bea
\label{final}
\int_{\mathbb{R}} d\tau\,g(t)^2\langle T^{ren}_{tt}\rangle_{\omega}
(t,0) \geq- \frac{1}{16\pi^2} \bigg\{& &I_1+\frac{5}{6}\Rmax J_2\\
&+&\Rmax''\left[\frac{23}{30}J_3+\left(\frac{43}{40}+16\pi^2(24|a|+11|b|)\right) J_4\right]
\nonumber\\
&+&\Rmax''' \left[\frac{163\pi+14}{96\pi}J_5
+\frac{7(2\pi+1)}{192\pi}(4J_6+J_7)\right] \bigg\}\,.\nonumber
\eea

Once we have a specific sampling function $g$, we can compute the
integrals of Eqs.~(\ref{J17}) to get a specific bound.  In the case of
a Gaussian sampling function,
\be\label{Gaussiang}
g(t)=e^{-t^2/t_0^2}\,,
\ee
we computed these integrals numerically in Ref.~\cite{Kontou:2014eka}.
Using those results the right hand side of Eq.~(\ref{final}) becomes
\be\label{finalGaussian}
-\frac{1}{16 \pi^2 t_0^3} \left\{ 3.76+2.63 \Rmax t_0^2+[3.42+ 197.9(24|a|+11|b|)] \Rmax'' t_0^4+ 6.99 \Rmax''' t_0^5 \right\}\,.
\ee
The leading term is just the flat spacetime bound of
Ref.~\cite{Fewster:1998pu} for $g$ given by Eq.~(\ref{Gaussiang}).
The possibility of curvature weakens the bound by introducing
additional terms, which have the same dependance on $t_0$ as in
Ref.~\cite{Kontou:2014eka}, with the Ricci tensor bounds in place of
the bounds on the potential.

\section{Conclusion}
\label{sec:conclusions}

In this work, using a general quantum inequality of Fewster and Smith
\cite{Fewster:2007rh} we derived an inequality for a minimally-coupled
quantum scalar field on spacetimes with small curvature. We calculated
the necessary Hadamard series terms and the Green's function for this
problem to first order in the curvature. Combining these terms gives
$\tilde{H}$ and taking the Fourier transform gives a bound in terms of
the Ricci tensor and its derivatives.

If we know the spacetime explicitly, Eqs.~(\ref{qinequality2}),
(\ref{BforR}), and (\ref{I}) give an explicit bound on the weighted
average of the energy density along the geodesic.  This bound depends
on integrals of the Ricci tensor and its derivatives combined with the
weighting function $g$.

If we do not know the spacetime explicitly but we know that the Ricci
tensor and its first 3 derivatives are bounded, Eqs.~(\ref{J17}) and
(\ref{final}) give a quantum inequality depending on the bounds and the
weighting function.  If we take a Gaussian weighting function,
Eq.~(\ref{finalGaussian}) gives a bound depending on the Ricci tensor
bounds and the width of the Gaussian, $t_0$.

As expected, the result shows that the corrections due to curvature
are small if the quantities $\Rmax t_0^2$, $\Rmax'' t_0^4$, and
$\Rmax''' t_0^5$ are all much less than 1.  That will be true if the
curvature is small when we consider its effect over a distance equal
to the characteristic sampling time $t_0$ (or equivalently if $t_0$ is
much smaller than any curvature radius), and if the scale of variation
of the curvature is also small compared to $t_0$.

In all bounds, there is unfortunately an ambiguity resulting from the
unknown coefficients of local curvature terms in the gravitational
Lagrangian.  This ambiguity is parametrized by the quantities $a$ and
$b$.

Ford and Roman \cite{Ford:1995wg} have argued that flat-space quantum
inequalities can be applied in curved spacetime, so long as the radius
of curvature is small as compared to the sampling time.  The present
paper explicitly confirms this claim and calculates the magnitude of
the deviation.  The curvature must be small not only on the path where
the quantum inequality is to be applied but also at any point that is
in both the causal future of some point of this path and the causal
past of another.  All such points are included in the integrals in
Eq.~(\ref{I5}) and (\ref{I6}).

Is is interesting to consider vacuum spacetimes, i.e., those whose
Ricci tensor vanishes.  These include, for example, the Schwarzschild
and Kerr spacetimes, and those consisting only of gravitational waves.
In such spacetimes, the flat-space quantum inequality will hold to
first order without modification.  There are, of course, second-order
corrections.  For the Schwarzschild spacetime, for example, these were
calculated explicitly by Visser
\cite{Visser:1996iw,Visser:1996iv,Visser:1997sd}.

In Ref.~\cite{Kontou:2012ve} we proved a theorem ruling out achronal
ANEC violation, given a conjecture that paths with small acceleration
in spacetimes with small curvature obey the same null-projected
timelike-averaged quantum inequality as in flat space
\cite{Fewster:2002ne}, with corrections of the form discussed here.
The present result is a step toward proving that conjecture.  In
future work we intend to extend the present result to null-projected
instead of timelike-projected quantum inequalities and to handle
slightly non-geodesic curves.

\section*{Acknowledgments}

We thank Larry Ford for helpful discussions.  This research was
supported by grant RFP3-1014 from The Foundational Questions Institute
(fqxi.org).  E.-A. K. gratefully acknowledges support from a John
F. Burlingame Graduate Fellowship in Physics."

\bibliography{no-slac,paper}

\end{document}